\documentclass[preprint,aps,showpacs]{revtex4}
\usepackage{graphicx}
\usepackage{dcolumn}
\usepackage{bm}
\usepackage{latexsym}
\usepackage{graphics}                               
\usepackage{amssymb} 
\input isolatin1.sty

\date{\today}
\begin{document}
\title{How the shift of the glass transition temperature
of thin polymer films depends on the adsorption with
the substrate
}
\author{Didier Long and
Paul Sotta}
\affiliation
{
Laboratoire de Physique des Solides, Universit\'e de Paris XI, Bât. 510,\\
91405 Orsay C{\'e}dex, France.\\
}
\begin{abstract} Recent experiments have demonstrated that
the glass transition temperature
of thin polymer films can be shifted as compared to the same
polymer in the bulk, the amplitude and the sign of
this effect depending on the interaction between the
polymer and the substrate.
A model has been proposed
recently for explaining these effects in two limiting
cases: suspended films and strongly adsorbed films. We extend
here this model for describing the cross-over between
these two situations. We show here how, by adjusting the strength of the adsorption,
one can control the glass transition temperature
of thin polymer films. In particular, we show that the shift of glass
transition temperature, refered to that of a suspended film,
varies like $\epsilon^{1/\gamma_2}$ where $\epsilon$ is the adsorption energy
per monomer and $\gamma_2$
is the critical exponent for the mass of aggregates in the 2D percolation problem.
We show also that the interaction leads to an increase of $T_g$ for $\epsilon$
as small as a few $0.01 T$ where $T$ is the thermal energy, and that this increase saturates 
at the value corresponding to strongly interacting films for adsorption energies 
of order $0.5 T$. 
\end{abstract}
\maketitle
\section{\textbf{Introduction}}
When cooling a glass forming liquid, instead of freezing at some
well defined temperature, one observes a huge increase of the
viscosity which takes place continuously. Such glass formers can
be either simple liquids or polymer liquids, and many features are
similar in both regarding the glass transition. One defines the
glass transition temperature $T_g$ as the temperature  at which
the dominant relaxation time on the molecular scale (or monomeric
scale in the case of polymers)
 reaches about $\sim 100 s$, which corresponds typically
to a viscosity of $10^{12} Pa.s$ in the case of simple liquids.
Typically for such glass forming liquids, the viscosity increases
by 12 orders of magnitude over a change of temperature of about
$100 K$ down to $T_g$. For reviews on the glass transition see
e.g. \cite{ediger,science}. The underlying mechanisms involved in
this dramatic increase are still poorly understood. One of the
issue has long been to know whether the relaxation mechanisms take
place on an increasing length scale when lowering the temperature
\cite{adams}. This issue lead to a huge interest in studying the
glass transition in thin liquid films over the past ten years
\cite{forrest3}. For practical reasons (film stability) these
studies have been performed on thin polymer films. Indeed it has
been demonstrated that the glass transition behaviour in thin
polymer films is very different from that of the bulk. When
considering a film deposited on a substrate with which the
interactions are supposed to be very weak, Jones et al
\cite{jones1} have shown that the glass transition temperature can
be reduced by as much as $25 K$ for $100 A$ thick films. Numerous
studies have confirmed this effect since then (see e.g.
\cite{for1,for2,jones2,mattsson,miyamoto}). In the case of polymers 
with radius of gyrations larger than the thickness of the film, 
Forrest and co-workers \cite{for1,for2} found even larger $T_g$ reductions, 
by as much as $70-80$ $K$ for $10$ $nm$ thick films. In this manuscript 
we will consider only the case of shorter chains, with radius of gyration 
smaller than the thickness of the film.  
On the contrary many
other  experiments performed on strongly adsorbed films have shown 
that the glass transition temperature increases in this case
\cite{wallace1,zanten1,grohens,grohens2,grohens3}, sometimes by as
much as $60K$ for films $80 A$ thick. More generally, it has been
demonstrated by various authors that both the sign and the
amplitude of the shift depends on the interaction with the
substrate \cite{nealey1,nealey2,russell,green,kremer2}. These
experiments were performed by measuring the thermal expansivity of
the films by ellipsometry
\cite{jones1,jones2,wallace1,zanten1,grohens,grohens2,grohens3,green}
or Brillouin light scattering \cite{for1}, by probing molecular
dynamics using dielectric spectroscopy \cite{kremer2,miyamoto} or
by calorimetric measurements \cite{nealey1,nealey2}.

Note that these $T_g$ shifts are very important effects if one
considers molecular relaxation times, or viscoelastic properties.
For instance -according to the usual WLF law- a shift of $T_g$ of a few tens of Kelvins results
 in an increase (or decrease) of
relaxation times by several orders of magnitude at a given
temperature $T$. The same polymer film is therefore expected to
have fundamentally different viscoelastic properties depending on
the substrate on which it is deposited. Note however that most
measurements of $T_g$ shifts have been performed either by
dilatometry (ellipsometry) or by dielectric measurements and are
somehow indirect. More direct measurements by Atomic Force
Microscopy have been reported e.g. in
\cite{hammerschmidt,tanaka,sokolov}, but brought conflicting
results. Indeed, some studies have confirmed previous results
\cite{hammerschmidt,tanaka} while another \cite{sokolov} failed to
reveal corresponding changes in the viscoelastic properties.
However, even though experimental studies are still required to
get a complete understanding of thin film behaviour, it is
generally considered now that huge dynamical changes take place in
thin films as compared to the bulk.

Another feature of glass transition which has been emphasized
recently is the strongly heterogeneous nature of the dynamics
close to $T_g$
\cite{spiess1,spiess2,ediger10,chamberlin,richert1,otpdeta,sillescu4,ediger1,
ediger9, ediger2,ediger3,ediger12,ediger4}. This heterogeneous
nature has been demonstrated by many experimental studies, using
techniques such as NMR \cite{spiess1,spiess2,ediger10},
fluorescence recovery after photo-bleaching (FRAP)
\cite{ediger1,ediger9, ediger2,ediger3,ediger12,ediger4},
translational and rotational probe diffusion
\cite{sillescu4,otpdeta,ediger9}, dielectric hole-burning
\cite{chamberlin} or solvation dynamics \cite{richert1}. These
studies have demonstrated the coexistence of domains with
relaxation time distributions spread over more than 3 to 4
decades at temperatures  typically 
$20$ $K$ above $T_g$. The characteristic size $\xi$ of these domains, measured
by NMR \cite{spiess2}, is typically 2 to 4 nm in the case of van
der Waals liquids and as small as 1 nm in glycerol
\cite{ediger10}.
 Using a deep photo-bleaching technique,
Ediger and co-workers have measured the life-time of these
heterogeneities, which they found comparable and even larger than
the dominant relaxation time $\tau_{\eta}$ \cite{ediger12} at
temperatures below $T_g$, while Spiess and coworkers
\cite{spiess2} found both times to be comparable. For reviews on
these issues, see e.g. \cite{sillescu,ediger5,richert}.

One of us recently proposed a model for the glass transition in
the bulk, which allowed both to interpret thin film experiments,
and to propose an explanation for the origin of the heterogeneous
nature of the dynamics in the bulk \cite{long,long2}, in the case
of van der Waals liquids. We proposed that the glass transition in
the bulk is controlled by the percolation of slow domains
corresponding to upwards density fluctuations on the scale of a
few nanometers. Note that the possible rôle of density
fluctuations had first been considered in references
\cite{moynihan,ediger8} and that the model proposed 
by Long et al can be considered as an extension of that 
proposed by Ediger. In the case of a suspended film, the
glass transition occurs when the slow domains percolate in the
direction parallel to the film, which requires a larger fraction
of slow domains and results in a decrease of the glass transition
temperature. In the case of films deposited on a substrate with
which the interactions are strong, the glass transition occurs
when both interfaces are connected with continuous paths of slow
domains. This requires that the correlation length of the 3D
percolation problem is comparable to the thickness of the film,
which requires a smaller fraction of slow subunits as compared to
the bulk glass transition. In both cases -suspended films and
strongly interacting films-  we predict a shift of the glass
transition of the form
\begin{equation}
\frac{\Delta T_g}{T_g} = \beta (\frac{a}{h})^{1/\nu_3}  \label{eq:1}
\end{equation}
where $\beta$ is a number of order unity. The sign
of $\beta$ is positive in the case of strongly interacting
films (SI films) and
negative in the case of suspended or weakly interacting films (WI films).
$\nu_3$ is the critical exponent for the correlation length in 3D percolation.
Note that the mechanism at work in the model proposed by Long and co-workers 
applies only for simple liquids, or short polymers, in that it does not 
predict molecular weight dependence of the $T_g$ shift such as those 
observed by Forrest and co-workers \cite{for1,for2}. Thereby we restrict our 
discussions below to the case of low molecular weight polymers, i.e. chains 
with radius of gyration smaller than the thickness of the film.   
Note that de Gennes \cite{degennes1} has proposed a mechanism at 
work in the case of large chains, which is not incompatible with 
the mechanisms described in \cite{long}. 

The aim of this paper is to describe the cross-over between the
two extreme cases, and thereby to give a more precise meaning to
the expressions``weakly interacting films'' or ``strongly
interacting films''. The issue is: what is the strength of the
polymer-substrate interaction required for observing an increase
of the glass transition? Are typical adsorption interactions
sufficient or is chemical grafting required? The paper is
organized as follows. First (section {\bf II}) we summarize
briefly the model for the glass transition in van der Waals
liquids proposed in \cite{long}. Then in section {\bf III} we
 re-discuss the glass transition in thin suspended films or in strongly interacting films.
 In section {\bf IV} we discuss the cross-over between
the WI case and the SI case precisely, and provide an expression
of the energy scale of the interaction at the cross-over. We discuss our 
predictions (section {\bf V}) and compare them 
to some recent experimental data
(section {\bf VI}). Note that in a joint paper \cite{sotta}, we study in more details
percolation properties in thin films in various situations
corresponding to the physical cases studied here, and we perform
Monte Carlo numerical simulations to illustrate and support our
theoretical considerations.

\section{\textbf{ Model for the glass transition in van der Waals liquid}}
As mentioned in the introduction, it has been demonstrated over
the past ten years that the dynamics is strongly heterogeneous
close to the glass transition \cite{sillescu,ediger5,richert}. One
of us \cite{long,long2} proposed that the slow subunits correspond
to upwards density fluctuations, and fast subunits to downwards
density fluctuations. We introduced a density $\rho_c$ above which
a microscopic domain of volume $v_0$ (to be defined below) is in a
very high viscosity state, i.e. has a relaxation time of order
$\tau_g$, where e.g. $\tau_g \sim 100s$ is the time scale chosen
for defining the glass transition. We proposed that the scale
$v_0$ is determined by two competing relaxation processes:
individual monomer jumps and the relaxation of the density of
dense subunits in the more mobile neighbouring subunits by a
diffusion process. The competition of these two processes results
in the building up of the slowest relaxation process at an
intermediate scale that we calculated to correspond to a number
$N_c$ of about a few hundreds of monomers \cite{long2}, which
corresponds to a characteristic length $\xi$ of 2 to 4 nanometers
and to a volume $v_0 \approx N_c a^3$ of order a few $10 nm^3$
typically ($a \approx 5 A$ is a typical monomer length), in
agreement with experimental results. The probability distribution
of the density of these small domains of volume $v_0$ is
\begin{equation}
P(T,\rho) \sim \exp(\frac{- (\rho_{eq}(T)-\rho)^2 v_0 K}{2 T\rho^2_{eq}}) \label{eq:2}
\end{equation}
where $\rho_{eq}(T)$ is the mean equilibrium density at
temperature $T$ and $K$ is the bulk modulus.
In our model, the glass transition is defined as
the temperature at which subunits of volume $v_0$ and
density larger than $\rho_c$ percolate.
According to our model, the heterogeneous nature of the dynamics results
from the strongly non-linear behavior of
the dynamics as a function of $\rho$ close to $T_g$.
More generally, at any temperature $T$,  
the viscosity is dominated by the characteristic time scale corresponding 
to the density $\rho_c(T)$, such that subunits of density equal or larger than 
$\rho_c(T)$ percolate, that is 
\begin{equation}
\int_{\rho_c(T)}^{\infty} P(T,\rho) d\rho = p^{3D}_c  \label{eq:3}
\end{equation}
which reads
\begin{equation}
\frac{\rho_c(T) -\rho_{eq}(T)}{\rho_{eq}} =  (\frac{T}{Kv_0})^{1/2} F(p^{3D}_c )   \label{eq:4}
\end{equation}
where $F(x)$ is the reciprocal function of the error function $Erf(x)$ defined by 
\begin{equation}
Erf(x) = \frac{1}{(2\pi)^{1/2}} \int_{x}^{+\infty} \exp(\frac{-u^2}{2}) du   \label{eq:5}
\end{equation}
Note that $F(x)$ is positive for $x<0.5$ and negative for $x>0.5$. $p^{3D}_c$ is 
a 3D percolation threshold. Thereby, the dominant relaxation time in a mechanical 
experiment is $\tau = \tau(\rho_c(T))$. It coexists with much faster relaxation times which are  
not relevant for the viscosity \cite{long,long2}. Knowing $N_c$, one can calculate $\rho_c-\rho_{eq}(T_g)$, 
using equation ~(\ref{eq:4}). 
If one considers for instance bond percolation on a FCC cubic
lattice, one has $p^{3D}_c \approx 0.119$. With the typical values
$N_c \approx 100$ and $K \sim 10^9$ Pa, one obtains that $(\rho_c
-\rho_{eq})/\rho_{eq}$ is of order a few percent.

\section{\textbf{Glass transition in thin suspended films and in films with strong interactions
with their substrate}}
\subsection{\textbf{Glass transition in thin suspended films}}
Consider first a film of small thickness $h$
of order a few hundred Angströms, either freely suspended or deposited on
a substrate with which interactions are weak.
In the same way as for the glass transition in the bulk, for the
thin polymer film to be in the glassy state, we assume that the
viscous domains of volume $v_0$  and density larger than $\rho_c$
percolate. The percolation that is required is in the direction
 of the plane parallel
to the film so as to build macroscopic clusters of slow dynamics.
On the scale $\xi = a N^{1/3}_c$, the film has a finite number $n
= h/\xi$ of layers in the direction normal to the film. For $\xi
\approx 2nm$ (which amounts to taking $N_c \approx 100$), $n$ is
of order a few tens. Then, as far as percolation is
concerned, the situation is intermediate between a 3D and a 2D problem: we
call it a quasi 2D problem. One can show \cite{sotta} that the
percolation threshold $p^+_c(n)$ in a film with a finite number
$n$ of layers is larger than the 3D percolation threshold and is
given asymptotically by
\begin{equation}
p^+_c(n) = p^{3D}_c + \mu n^{-1/\nu_3}    \label{eq:6}
\end{equation}
where $\mu$ is a number of order unity ($\approx 0.5$ for a cubic
lattice), and $\nu_3 \approx 0.88$ is the critical exponent of the
correlation length in 3D percolation. Thus, the cross-over between
2D and 3D is described by a power law and is wide. For the film to be in
the glassy state, one needs a fraction $p^+_c(n)$ of slow
subunits, which is larger than that required in the bulk,
$p^{3D}_c$. By calculations similar as those used to calculate
$\rho_c$, it has been shown in \cite{long} that the relation between the temperature $T$ and the
fraction $p$ of subunits of volume $v_0$ with density larger than
$\rho_c$ is given by 
\begin{equation}
\frac{T- T_g}{T_g} = (F(p)-F(p^{3D}_c)) \frac{1}{N_c^{1/2}}      \label{eq:7}
\end{equation}
where we have used the relations $(K/(T\rho_{eq}))^{1/2}\chi_T \approx 1$
($\chi_T$ is the dimensionless thermal expansion coefficient) \cite{long}.
Since we consider only small differences for $p-p^{3D}_c$, this equation can be linearised
to yield
\begin{equation}
p - p^{3D}_c = \frac{T-T_g}{ T_g} \frac{N_c^{1/2}}{F'(p^{3D}_c)} = -\gamma \frac{T-T_g}{ T_g}
\label{eq:8}
\end{equation}
with $F'(p) = d F/dp$ and $\gamma = - N_c^{1/2}/F'(p^{3D}_c)>0$. 
Writing equation~(\ref{eq:8}) with $p = p^+_c(n)$, and using
equation~(\ref{eq:6}), one obtains the glass transition
temperature $T^-_g(n)$ of the film:
\begin{equation}
\frac{\Delta T^-_g(n)}{T_g} = \frac{-1}{\gamma} (p^+_c(n) - p^{3D}_c) =
\frac{-\mu}{\gamma} n^{-1/\nu_3} =  \frac{-\mu}{\gamma} (\frac{h}{\xi})^{-1/\nu_3}  \label{eq:9}
\end{equation}
Note that we are using the notation $p^+_c(n)$ here because
$p^+_c(n)$ is larger than $p^{3D}_c$ and $\Delta T^-_g(n)$ because
$T_g$ is shifted downwards in this case ($\Delta T^-_g(n)<0$), as
it is clear in equation~(\ref{eq:9}). From this result, one
deduces the $T_g$ shift:
\begin{equation}
\frac{\Delta T^-_g(h)}{T_g} = \frac{\mu F'(p^{3D}_c)}{N_c^{(1/2-1/(3\nu_3))}} (\frac{a}{h})^{1/\nu_3}
= -\beta(\frac{a}{h})^{1/\nu_3}   \label{eq:10}
\end{equation}
which we can write also
\begin{equation}
T^-_g(h)
= T_g(1-\beta(\frac{a}{h})^{1/\nu_3})   \label{eq:10b}
\end{equation}
with $\beta = -\mu F'(p^{3D}_c)/N_c^{(1/2-1/(3\nu))}$ and $a$
is the monomer length (typically $5 A$). With $N_c \sim 100$, $\nu \approx 0.88$,
$\mu \approx 0.5$ \cite{sotta},  $p^{3D}_c =0.119$ \cite{stauffer} and $F'(0.119) \approx -4$, one
obtains that $\beta \approx 1.5$. In the case of polystyrene, for which $T_g \approx 375 K$, 
this leads to a decrease of $T_g$ of about $20 K$ for a film $10$ $nm$ thick, which 
is in agreement with typical experimental results \cite{jones1,jones2}. 
One of the most important conclusion here is that the $T_g$ shift 
that we predict is long ranged. According to our model,
the $T_g$ shifts measured e.g. in ellipsometric experiments
such as in references \cite{jones1,for2}
do not result from an average effect of strongly perturbed
thin layers at the interfaces of the film: the ``bulk'' of the
film itself has a lower $T_g$, as given by equation ~(\ref{eq:10b}).

Fig.1 illustrates our model for the glass transition in thin
suspended films. One can see from above a spherical body, with a
radius larger than the thickness $h$ of the film. Both caps of the
particle emerge from the film on each side. At a temperature
higher than $T_g+\Delta T^-_g(h)$ the slow aggregates of
relaxation time $\tau_g$ may connect both surfaces, but they do
not percolate in the direction parallel to the film. They build
only finite clusters, which move freely in the film. The viscosity
of the film is measured by pulling the spherical body. It is
smaller here than the viscosity in the bulk and it is dominated by
the slowest subunits which percolate in the direction parallel to
the film. The measured viscosity $\eta_{film}$ of the film may be
calculated from the $T_g$ shift that we calculated before  :
\begin{equation}
\eta_{film} (T,h) = \eta_{bulk}(T-\Delta T^-_g(h))   \label{eq:11}
\end{equation}
where $\Delta T^-_g(h)$ is given by equation ~(\ref{eq:10}) and
$\eta_{bulk}(T)$ is given by the corresponding WLF (or VFT) law of
the considered liquid (remember that $T-\Delta T^-_g(h)$ is higher
than $T$ since $\Delta T^-_g(h)$ is negative). The viscosity which
dominates the motion of the body is that in the middle of the
film, where it is the highest. Note that in a two layer model, in which 
the measured $T_g$ shift would result from an average
between a fluid thin layer and an unperturbed ``bulk'' of the
film, the viscosity of the film would be given by the viscosity of
the bulk. Thereby, an experiment analogous to that proposed here (cf Fig.1), or an analogous 
one, is discriminating between our interpretation and that 
of the two layer model \cite{for2,jones2}. At the
temperature $T_g+\Delta T^-_g(h)$ (Fig. 1.b.) the slow subunits
percolate in the direction parallel to the film and the film is at
its glass transition: in order for the body to move, the slow
subunits have to deform and reorganise.  The viscosity of the
film measured by pulling the body is thereby the same as that of
the bulk at $T_g$.

\subsection{\textbf{Glass transition in films with strong interactions with their substrates}}
Let us consider now a thin polymer film with strong interactions
with the substrate, at a temperature $T>T_g(bulk)$. The
correlation length of slow subunits, i.e. the size of slow
aggregates is $\zeta \sim \xi |p-p^{3D}_c|^{-\nu_3}$ -where $p$ is
the fraction of slow domains at temperature $T$. If $p$ is
sufficiently close to $p^{3D}_c$, i.e. if the temperature is not
too far above $T_g$, the size of slow aggregates is comparable to
the thickness $h$ of the film. This means that there is a finite
fraction of monomers from the free interface which are connected
to the rigid substrate by continuous path of slow subunits. Then,
the film is glassy. Indeed, consider a shear experiment between
the two interfaces: there is a no-slip boundary condition on the
upper interface, while the monomers are attached to the other: one
needs therefore to deform slow aggregates which thereby control
the viscosity of the film. The condition for being glassy 
reads then:
\begin{equation}
\zeta \sim \xi |p-p^{3D}_c|^{-\nu_3} \sim h  \label{eq:12}
\end{equation}
Thereby a fraction of slow subunits $p^-_c(n)$ smaller than
$p^{3D}_c$ is required at the glass transition. It is given by
\begin{equation}
p^{3D}_c - p^-_c(n) \approx  n^{-1/\nu_3}    \label{eq:12b}
\end{equation}
which results in an increase of $T_g$ given by
\begin{equation}
\frac{\Delta T^+_g(h)}{T_g} = \beta'(\frac{a}{h})^{1/\nu_3}   \label{eq:13}
\end{equation}
which we can write also:
\begin{equation}
T^+_g(h) =T_g (1+ \beta'(\frac{a}{h})^{1/\nu_3})   \label{eq:13b}
\end{equation}
where $\beta' \approx -F'(p^{3D}_c)/N_c^{(1/2-1/(3\nu_3))}$ is
analogous to $\beta$, with a different pre-factor however. To
illustrate our model, one may also adopt an equivalent point of
view, by considering a semi-infinite polymer melt in contact with
a strongly interacting substrate, at a temperature $T > T_g$ where
$T_g$ is the glass transition temperature in the bulk. Monomers
directly in contact with the substrate are immobilized. Then, one
may ask: is the melt in a glassy state in the vicinity of the
substrate? If yes, what is the range of this effect? We argue that
at a distance $z$ smaller or comparable to the 3-D percolation
correlation length $\zeta \sim \xi |p-p^{3D}_c|^{-\nu_3}$ -where
$p$ is the fraction of slow domains at the temperature $T$-, a
finite fraction of monomers are connected to the substrate by
continuous paths of slow dynamics domains. Then, these monomers
will control the dynamics at the distance $z$, as a consequence of
the strong adsorption. Indeed, without adsorption, the continuous
path may move freely within the melt since $T > T_g$. Then, the
polymeric liquid is glassy within a distance $d$ to the substrate
given by
\begin{equation}
d =  b' (\frac{T_g}{T - T_g})^{\nu_3}   \label{eq:14}
\end{equation}
with $b' = a\beta'^{\nu_3}$ is of order one monomer length, i.e.
$5 A$. Note that at about $T = T_g +10K$, the distance $d$ can be
a few tens of nanometers. We have just emphasized that the effects
we predict are long ranged. Indeed our model predicts a shift of
$T_g$ given by a power law as a
function of the distance to the interface. In particular, we argue
against the interpretation proposed by some authors (see e.g.
\cite{mattsson}) according to which the measured $T_g$ shifts result from an
average effect, between strongly perturbed thin interfacial layers
and an essentially unperturbed ``bulk'' of thin films. This point
is essential when one considers viscoelastic properties of thin
films \cite{hammerschmidt,tanaka,sokolov} or of filled materials
\cite{landel,edwards,berriot}.
\\

The case of a SI film is schematised on Fig. 2. Let us consider
here a body, such as the tip of an AFM, penetrating in the film,
within a distance $z$ to the substrate. We suppose that the
interactions between the tip and the polymeric liquid are weak, whereas they
are strong between the liquid and the substrate. There is indeed a
no slip boundary condition on the surface of the tip, whereas the
monomers in the aggregates are attached to the substrate. At a
temperature $T$ larger than $T_g$, and at a distance $z$ larger
than the size $\zeta$ of the slow aggregates, the viscosity
controlling the motion of the AFM tip is smaller than that at
$T_g$. Indeed, at this distance, the slow aggregates move freely
in the film and can move away rigidly to open a path for the tip.
When the tip gets closer to the substrate, the viscosity increases
(Fig. 2.b). Indeed, at distances smaller than $\zeta$, the AFM tip
has to deform the slow aggregates in order to move parallel to the
film. More precisely, the viscosity measured by the probe is then:
\begin{equation}
\eta_{film} (T,z) = \eta_{bulk}(T-\Delta T^+_g(z))   \label{eq:15}
\end{equation}
where $\Delta T^+_g(z)$ is given by equation ~(\ref{eq:13}) and
$z$ is the distance to the substrate (here $\Delta T^+_g>0$, so
that $T-\Delta T^+_g<T$). Note the difference between the no slip
boundary condition at the surface of the tip (if interactions
between the tip and the monomers are weak) and the fact that the
monomers on the substrate are attached. That latter condition is
much more restrictive. A no slip boundary condition allows the
monomers in contact with the tip to be renewed, which is not the
case for the monomers in contact with the substrate if they are
attached. For instance, if there was no specific interaction
between the monomers and the substrate, one would have a no slip
boundary condition also at the film-substrate interface. Then the
slow aggregates encountered by the tip (Fig. 2.b) could rotate and
reorganise within the film.
 According to our model, this difference results in the huge 
difference in viscosity between the cases of suspended
films and SI films, as illustrated by equations ~(\ref{eq:11}) and
~(\ref{eq:15}). More precisely, the viscosity of the film can be
calculated using equation ~(\ref{eq:15}) and the bulk WLF law (or
VFT law in the context of simple liquids) \cite{ferry,handbook}:
\begin{equation}
\log(\frac{\eta_{bulk}(T)}{\eta_{bulk}(T_{g})}) = \frac{-C_1 (T-T_{g})}{C_2+T-T_g}
                                     \label{eq:15b}
\end{equation}
 $C_1$ and $C_2$ are two
constants which depend on the considered polymer. The viscosity at
$T_g$, $\eta(T_g)$ is typically of order $10^{12}$ $Pa.s$ in the
case of simple liquids. In the case of polymeric liquids, it is more 
appropriately defined a monomeric jump time of order $100$ $s$ 
\cite{ferry,handbook}. Close to $T_g$, the slope of $\log(\eta(T))$ is very
large, such that an increase of $T_g$ of $20$ $K$ results in an
increase of the viscosity up to 3 or 4 orders of magnitude
depending of the liquids. When compared to the opposite case of a
WI film, the same polymer films at the same temperature can have
viscosities which differ by more than 6 or 7 orders of magnitude
depending on their substrates.

\section{\textbf{Adsorbed polymer film}}
We have just seen that when the film-substrate interaction is
strong, the film is glassy when the size of the aggregates of slow
subunits is comparable to the thickness of the film. This occurs
when the correlation length of the 3D percolation problem is
comparable to the thickness $h$ of the film. The main point is
that the monomers in contact with the substrate are ``attached''.
As a consequence, a probe which moves within the film deforms slow
aggregates, which thereby control the viscosity of the film.
However, when the interactions with the substrate are finite, a monomer remains 
anchored on the substrate during a finite time. Thus the issue is: 
which interaction strength is required for the film to be glassy, or can
we calculate the $T_g$ shift as a
function of the interaction strength? Are typical
adsorption interactions sufficient for inducing an increase of $T_g$?
 Let us define $\epsilon$ as the
adsorption energy of one monomer with the substrate. If one could
remove a monomer independently of the other ones, the life-time
$\tau_d$ of monomer adsorption on one site would be:
\begin{equation}
\tau_d = \tau_0\exp({\frac{\epsilon}{T}})    \label{eq:16}
\end{equation}
where $\tau_0$ is a microscopic time of typical order of magnitude
$10^{-12}$ $s$. For the film to be in the glassy state, the time
$\tau_d$ must be at least equal to $\tau_g \sim 100$ $s$. However,
typical adsorption energies are of order a few $T$ at most, which
is far from required.

However, one must consider the number of monomers which must be
removed {\em collectively on the time scale $\tau_g$}. These are
the ensemble of monomers which belong to the same aggregate of
slow subunits and which are in contact with the substrate. One can
see such an aggregate on Fig. 4. We call $S = S(h,p)$ the number
of such monomers. On a time scale smaller than $\tau_g$, such an
aggregate must be considered to be rigid. The time scale for
removing such an aggregate from the substrate is therefore
\begin{equation}
\tau_d = \tau_0\exp({\frac{f(\epsilon)\epsilon S}{T}})  \label{eq:17}
\end{equation}
where $f(\epsilon)$ is the fraction of monomers in contact with
the substrate which are adsorbed. In a first approach, this
function can be considered to be of order unity and shall be omitted in
the following. Of course equation~(\ref{eq:17}) holds as long as 
the r.h.s is smaller than $\tau_g$, since the aggregate
does not persist on longer time scales. This equation breaks down
at low $\epsilon$ when the quantity $\epsilon S/T$ is smaller than
one. Then the time for removing the aggregates must be considered to be 
equal to $\tau_0$.
 Now, when considering
equation ~(\ref{eq:17}), one can understand why typical adsorption
energies are sufficient for increasing the glass transition temperature
of thin films. Indeed, when the size of the aggregates is comparable
to the thickness $h$ of the film, the number $S$ is at least equal
to $N^{2/3}_c$, which is the number of monomers on one interface of
a slow subunit. Then, this number is of order a few tens, which can be
sufficient for $\tau_d$ being comparable to $\tau_g$, depending
on the adsorption energy $\epsilon$.

Let us describe now in more details the glass transition mechanism that we propose
for thin films deposited on a substrate with interaction energy $\epsilon$
per monomer. First let us note that the relevant temperature range is
\begin{equation}
T^-_g(h) < T < T^+_g(h)  \label{eq:18}
\end{equation}
where $T^-_g(h)$ and $T^+_g(h)$ are the glass transition temperatures 
of a freely suspended film and of a strongly adsorbed film and 
are given by equations (11) and (16) respectively. 
Indeed for $T>T^+_g(h)$ a film is fluid whatever the interactions
with the substrate according to previous discussions. For
$T<T^-_g(h)$ a film is glassy, even if the interactions with the
substrate vanish, as in the case of suspended films. In terms of
the fraction of slow aggregates, this temperature range
corresponds to:
\begin{equation}
p^+_c(n) > p > p^-_c(n)   \label{eq:19}
\end{equation}
where $p^+_c(n)$ is given by equation ~(\ref{eq:6}) ($n = h/\xi$) and is larger
than $p^{3D}_c$, and $p^-_c(n)$ is smaller than $p_c^{3D}$ and is
given by equation ~(\ref{eq:12b}). Both curves $p^+_c(h)$ and
$p^-_c(h)$ are shown in  Fig. 3. Both quantities correspond to a
correlation length for the 3D percolation problem comparable to
the thickness $h$ of the film. One ($p^-_c(h)$) is below the
percolation threshold (case of strongly adsorbed film), the
other one ($p^+_c(h)$) is above the percolation threshold (case of a
suspended film).

Let us start from the fluid state at $T>T^+_g(h)$. By lowering the
temperature below $T^+_g(h)$ (i.e. by increasing $p$ above
$p^-_c(n)$), the size of the aggregates increases and they attain a
lateral extension $\zeta_{\|}$ larger than the thickness $h$ of
the film: one enters the cross-over between 2D and 3D percolation.
One can see such an aggregate on Fig. 4. We show in \cite{sotta}
that this quasi-2D percolation problem with probability of
occupation $p$ and mesh size $\xi$ can be mapped, using a
coarse-graining function $\Phi$, into a 2D problem of mesh size
$h$ and probability of site occupation $\Phi^{(N)}(p)$ with
$\Phi^{(N)}(p^+_c(h)) = p_c^{2D}$. $N$ is the number of iterations
for transforming the film of thickness $n$ on the scale $\xi$ ($h
= n \xi$) into a 2D lattice of mesh size $h$. Using this
procedure, one can demonstrate  that the mass $M'$ of an aggregate
is \cite{sotta}:
\begin{equation}
M'(h,p) \approx |\Phi^{(N)}(p^+_c(n)) -\Phi^{(N)}(p)|^{-\gamma_2}
\approx (\frac{h}{\xi})^{-\gamma_2/\nu_3} |p^+_c(n) -p|^{-\gamma_2}  \label{eq:20}
\end{equation}
where the mass $M'$ is given in unit of cubes of size $h^3$ in the
renormalised system. On the other hand, the correlation length
(expressed in the renormalised system, i.e. in units of $h$) is:
\begin{equation}
\zeta'_{\|} (h,p) \approx |\Phi^{(N)}(p^+_c(n)) -\Phi^{(N)}(p)|^{-\nu_2} \approx
(\frac{h}{\xi})^{-\nu_2/\nu_3} |p^+_c(n) -p|^{-\nu_2}  \label{eq:21}
\end{equation}
where $p^+_c(n)$ as a function of $n$ is given in
equation~(\ref{eq:6}). All these relations hold within a
pre-factor of order unity which can be measured by numerical
simulations \cite{sotta}. The exponents $\nu_2 \approx 1.333$ and
$\gamma_2 \approx 2.389$ are the 2D critical exponents for the
correlation length and for the mass of the aggregates,
respectively. In units of mesh size $\xi$, the mass reads
\begin{equation}
M(h,p) \approx M'(h,p) (\frac{h}{\xi})^{2-\eta_3}   \label{eq:22}
\end{equation}
where $\eta_3 \approx -0.068$ is the anomalous exponent of the 3D
correlation function of the percolation problem. In physical unit
length, the correlation length parallel to the film reads
\begin{equation}
\zeta_{\|} \approx h (\frac{h}{\xi})^{-\nu_2/\nu_3} |p^+_c(n) -p|^{-\nu_2}   \label{eq:23}
\end{equation}
Note that the above result insures that for $p = p^-_c(n)$, i.e.
when the $3D$ correlation length is equal to the thickness of the
film, the size of the aggregates is equal to $h$ as it should be. 
From equations ~(\ref{eq:20}) and ~(\ref{eq:22}), one deduces that
the number of monomers from one aggregate in contact with the
substrate
 is (within a pre-factor of order unity)
\begin{equation}
S(h,p) \approx N_2 M(h,p) (\frac{h}{\xi})^{-1} \approx
N_2 (\frac{h}{\xi})^{1-\eta_3} (\frac{h}{\xi})^{-\gamma_2/\nu_3} |p^+_c(n) -p|^{-\gamma_2}
\label{eq:24}
\end{equation}
where $N_2 \approx N^{2/3}_c$ is the number of monomers on the
interface of a subunit of size $\xi$. Equations ~(\ref{eq:20}),
~(\ref{eq:22}-\ref{eq:24}), together with equation ~(\ref{eq:8})
which gives the variation of $p$ as a function of the temperature,
and ~(\ref{eq:6}) giving $p^+_c(n)$, provide the variation of the
mass of the aggregates, of the correlation length parallel to the
film, and of the number of monomers from one aggregate in contact
with the substrate, as a function of the temperature, for $T <
T^+_g$. At $T=T^-_g$, all these quantities diverge. From equations
~(\ref{eq:17}) and ~(\ref{eq:24}), one can calculate the time for
removing an aggregate from the substrate. At the glass transition
temperature
 of the film, this time must be equal to $\tau_g$, which reads
\begin{equation}
\tau_d \approx \tau_0 \exp(\frac{\epsilon S(h,p)}{T}) = \tau_g     \label{eq:25}
\end{equation}
This yields
\begin{equation}
\frac{\epsilon}{T} = (N_2 (\frac{h}{\xi})^{1-\eta_3})^{-1}
(\frac{h}{\xi})^{(\gamma_2/\nu_3)} \ln(\frac{\tau_g}{\tau_0})
|p^+_c(n)-p|^{\gamma_2}                                         \label{eq:26}
\end{equation}
This equation gives the fraction $p$ of slow subunits in the film
at the glass transition as a function of the interaction energy
$\epsilon$. The glass transition temperature $T$ of the film is
then obtained using equation~(\ref{eq:8}) which relates the
fraction $p$ of slow subunits and the temperature: $p - p^{3D}_c =
-\gamma (T-T_g)/T_g = -\gamma \Delta T_g/T_g$ , where
$\gamma=N_{c}^{1/2}/F'(p_{c}^{3D})$ is positive. Using the
relation $p^+_c(n) - p^{3D}_c = -\gamma \Delta T^-_g/T_g$, where
$\Delta T^-_g <0$ is the $T_g$ shift of thin suspended films, one
obtains
\begin{equation}
\frac{\Delta T_g}{T_g} = \frac{1}{\gamma} (\frac{\epsilon  N_2 (\frac{h}{\xi})^{1-\eta_3}}
{T \ln(\tau_g/\tau_0)})^{1/\gamma_2} (\frac{h}{\xi})^{-1/\nu_3} + \frac{\Delta T^-_g}{T_g}
\label{eq:27}
\end{equation}
or, using equation ~(\ref{eq:10}):
\begin{equation}
\frac{\Delta T_g}{T_g} = \frac{1}{\gamma}(\frac{h}{\xi})^{-1/\nu_3}
 ((\frac{\epsilon N_2 (\frac{h}{\xi})^{1-\eta_3}}
{T \ln(\tau_g/\tau_0)})^{1/\gamma_2} - \mu)         \label{eq:28}
\end{equation}
This equation describes the cross-over between a suspended film
and a film with strong interaction with its substrate, as a
function of the adsorption energy per monomer $\epsilon$. At low
values of $\epsilon$, one recovers the $T_g$ shift of suspended
films, by setting $\epsilon = 0$ in the above equation. For strong
interactions, the $T_g$ shift given by equation ~(\ref{eq:28}) is valid as long as 
it is smaller than $\Delta T^+_g$, i.e. as
long as the correlation length is larger than $h$. Equation
~(\ref{eq:28})  must be considered as leveling off at $\Delta
T^+_g$ when increasing $\epsilon$. Indeed, consider a system for
which $\epsilon$ can be tuned. At $\epsilon = 0$, the shift of
$T_g$ is negative. The lateral size of the aggregates is infinite.
When increasing $\epsilon$, the glass transition temperature
increases. At $\epsilon = \epsilon_c$, the shift cancels. At
larger value, the shift of $T_g$ is positive. The lateral size of
the slow aggregates has decreased but is still so large that the
number of monomers of one aggregate in contact with the substrate
is large enough for the desorption time of the aggregate to be
equal to $\tau_g$. In this regime the lateral size $\zeta_{\|}$ is
still larger than the thickness $h$ of the film. At some value
$\epsilon = \epsilon_{max}$, the glass transition temperature
corresponds to a 3D size of the aggregates comparable to $h$.
Then, increasing further the adsorption energy does not increase
further the glass transition temperature since it would correspond 
to a correlation length much smaller than the thickness of the film. 
This plateau value is
reached when $\epsilon$ is a few $\epsilon_c$  as given below, and
the corresponding $T_g$ shift is equal to $\Delta T^+_g$.
 In any case our model predicts:
\begin{equation}
\Delta T^-_g(h) \le \Delta T_g \le \Delta T^+_g(h)              \label{eq:29}
\end{equation}
The interaction can be considered to be strong when
\begin{equation}
 \frac{\epsilon   N_2 (\frac{h}{\xi})^{1-\eta_3}}
{T \ln(\tau_g/\tau_0)} > 1                                \label{eq:30}
\end{equation}
The interaction is weak, and the situation of the film is equivalent to that
of a suspended film when
\begin{equation}
 (\frac{\epsilon   N_2 (\frac{h}{\xi})^{1-\eta_3}}
{T \ln(\tau_g/\tau_0)})^{1/\gamma_2} \ll 1                \label{eq:31}
\end{equation}
Thereby, the cross-over between a freely suspended and a strongly
interacting film of thickness $h$ is given by
\begin{equation}
\epsilon_c \approx \mu^{\gamma_2} T
\frac{\ln(\tau_g/\tau_0)} {N_2 (\frac{h}{\xi})^{1-\eta_3}}  \label{eq:32}
\end{equation}
With $N_c$  $\sim 100$,
 $N_2 \sim N^{2/3}_c$ is of order a few tens.
The time scale chosen for the glass transition is $\tau_g \approx
100s$, which means that the quantity $\ln(\tau_g/\tau_0) \sim 30$
typically. With $h/\xi = 10$ for instance and $\mu \sim 0.5$, one
obtains then
\begin{equation}
\epsilon_c \sim 0.02 T                                      \label{eq:33}
\end{equation}
This value for the interaction per monomer can be considered to be
typical at the cross-over between negative shift of $T_g$ and
positive shift of $T_g$. Expressed in terms of the interaction
energy between the substrate and the polymer melt, one obtains
$\epsilon_c \sim 0.4$ mJ/m$^2$. Note that the thicker is the film,
the sharper is the transition between the two regimes (WI films
and SI films), but the smaller ($\propto (h/\xi)^{-1/\nu_3}$) the
amplitude of the effect. One can write equation ~(\ref{eq:28}) as:
\begin{equation}
\frac{T_g(h)-T^-_g(h)}{T_g} = \frac{1}{\gamma}(\frac{h}{\xi})^{-1/\nu_3}
 (\frac{\epsilon}{\epsilon_c})^{1/\gamma_2} \propto \epsilon^{1/\gamma_2}    \label{eq:35}
\end{equation}
to emphasize that $(T_g(h)-T^-_g(h))/T_g$ increases like
$\epsilon^{1/\gamma_2} \approx \epsilon^{0.418}$ in the crossover
region, a relation which may perhaps be tested experimentally. The
glass transition temperature as a function of the interaction per
monomer $\epsilon$ is plotted on Fig. 5. Finally, the $T_g$ shift
can be written as a function of the thickness of the film and of
the interaction energy as:
\begin{equation}
\frac{\Delta T_{g}(h,\epsilon)}{T_{g}}=\frac{\Delta T_{g}^{-}(h)}{T_{g}}
\left[ 1- \left(\frac{\epsilon}{\epsilon_c} \right)^{1/\gamma_{2}}
\right]
                                                 \label{eq:36}
\end{equation}
in which $\Delta T_{g}^{-}(h)$ is the negative shift in a
non-adsorbed film (cf equation ~(\ref{eq:10b})). The crossover value $\epsilon_c$ contains
itself a $h$ dependence and is given by equation ~(\ref{eq:32}). Equation
~(\ref{eq:36}), together with equations ~(\ref{eq:10b})  and ~(\ref{eq:32}),
is the main result of this manuscript.

As already discussed, equation ~(\ref{eq:36}) levels off at
$\epsilon = \epsilon_{max}$ so that the $T_g$ shift is equal to
$\Delta T^+_g(h)$ in the regime of strong adsorption. One has thereby 
\begin{equation}
\epsilon_{max}(h) = (1-\frac{\Delta T^+_g(h)}{\Delta T^-_g(h)})^{\gamma_2} \epsilon_c(h) 
= (1+\frac{\beta'}{\beta})^{\gamma_2}\epsilon_c(h)
\end{equation}
The ratio $\beta'/\beta$ is the ratio of the $T_g$ shift between 
the case of suspended films and that of strongly adsorbed films. Experimental 
data show that the typical shifts for $10$ $nm$ thick films are $20$ $K$ and 
$50$ $K$ in these respective cases. With $\gamma_2 \approx 2.389$ \cite{stauffer}, 
one obtains $\epsilon_{max} \approx 20 \epsilon_c$, or given our prediction 
for $\epsilon_c$ for a film $10$ $nm$ thick, we obtain 
\begin{equation}
\epsilon_{max}(h = 10 nm) \sim 0.5 T
\end{equation}
which, expressed in adsorption energy per unit area, corresponds to 
about $\sim 10 mJ/m^2$. Considering the respective values 
of $\epsilon_c$ and $\epsilon_{max}$, one sees indeed that observing a negative shift 
of $T_g$ in a film deposited on a rigid substrate requires very well 
controlled surfaces, with very low interactions between the substrate 
and the monomers.

\section{\textbf{Discussion}}
At this point, let us discuss some important aspects of the glass 
transition in thin films that can be deduced from our 
model. 
First, let us emphasise here that our assumption is that 
the local dynamics on small scale is the same in the bulk and in thin films. 
The change of dynamical properties between both situations take place 
on larger scale due to percolation mechanism, which differ between 
the bulk and thin films. This assumption is underpined by the fact that 
the statistics are the same between both cases: the perturbations of the 
thermodynamics due to the presence of interfaces is short ranged \cite{long} and 
do not extend beyond $1$ or $2$ $nm$. 

An important consequence of our model  is that the measured shift of $T_g$ should depend 
on the type of technique used for measuring the dynamics. This a 
consequence of the heterogeneous nature of the dynamics. The modification 
of the dynamics in thin films as respect to the bulk might be different for 
instance when considering viscoelastic properties and the 
diffusion of probes, as it has already been observed in the bulk, 
as a function of temperature \cite{otpdeta,sillescu4,ediger1,ediger9}. 
Note that the size of the probe plays a crucial rôle at this respect \cite{long2}.  
We expect also that the shift measured e.g. by NMR experiments, 
if any, should be different as  that measured by mechanical experiments. 
Indeed, in a mechanical experiment as described in this manuscript, one probes only 
a part of the dynamical spectrum (as defined by equation ~(\ref{eq:3})), 
according to the model proposed in \cite{long,long2}, whereas a NMR experiments 
probes a more local environment. 

  However, the $T_g$ shift should also depend, for a given technique, 
of some precise aspects of the experiment. For instance, we argued 
that a suspended film is more fluid when considering pulling a probe 
in the film, while for the same experiment we expect a strongly 
adsorbed film to be more viscous. Consider now an experiment in which 
one shears the film between two rigid substrates, which have no interaction 
with the considered liquid. Then, the $T_g$ shift should be different 
from that measured by pulling the probe and discussed in this paper. 
Indeed even at temperatures above $T^-_g$, the slow 
aggregates are strongly anisotrope and are very elongated in the 
direction parallel to the film, if $T$ is not too far above $T^-_g$. 
Then, in a shearing experiment as discussed here, 
one needs to deform them given the no-slip boundary condition. 
Some dilatancy is also expected, depending 
on the shearing rate. In other words, we expect the viscosity of the film 
to be strongly anisotropic.

Another issue is a more precise calculation of $\Delta T^+_g$. 
Indeed, our result here (equation ~(\ref{eq:13})) has been obtained 
only within a pre-factor, assuming the relation  
$p^{3D}_c -p^-_c \approx (h/\xi)^{-1/\nu}$, so that a fraction of order one 
of slow subunits on one interface are connected to the other interface by a slow 
aggregate. But we did not specify the value of this fraction.  
Thereby, the pre-factor $\beta'$ in equation ~(\ref{eq:13}) must be 
considered to be an adjustable parameter, which we choose to be  
consistent with a $T_g$ shift of about $+50$K for 
strongly interacting films of
thickness $10$nm, which is close to experimental results
\cite{wallace1,grohens,grohens2}. 
For the type of experiments discussed here (e.g. probe motion, Fig. 1 and Fig. 2), the viscosity 
of the suspended film is controlled by the slowest percolating subunits, leading 
to an equation analogous to equation ~(\ref{eq:3}) with $p^+_c(n)$ instead 
of $p^{3D}_c$. The percolation threshold in the direction parallel to the film 
is a well defined state, which is also the case in the bulk. It allows 
thereby to calculate the shift of $T_g$ between 
suspended films and the bulk as corresponding to the shift 
of the percolation threshold. In particular  
we expect the glass transition in thin 
suspended film as discussed here to be similar to that in the bulk. 
The situation of strongly adsorbed film is very different. The fraction of 
slow subunits at the free interface connected to the lower interface is never 
zero, but decays exponentially with the thickness $h$ of the film: the percolation 
between both interfaces is not a well defined state, but occurs continuously. 
In addition to the broad distribution of relaxation times already present in the 
bulk and in thin films, this fact tends to increase the width of the glass transition, 
as compared to the bulk 
or as compared to suspended films. This fact might be observed experimentally. 
For our purpose here, it makes a calculation of the $T_g$ shift more involved, as 
it requires a more precise model as that presented here 
for calculating the viscosity of the film as a function 
of the statistics of the slow aggregates.  
 

\section{\textbf{Comparison with experimental results}}

We have described in this paper a mechanism which leads to a $T_g$
shift in thin polymer films. This shift depends on the film
thickness and it can be either positive or negative depending on
the interaction between the monomers and the substrate. Two
limiting cases must be considered when dealing with equation
~(\ref{eq:36}): 1) the case of very weak interactions. Then, the
$T_g$ shift is negative and equal to $\Delta T_{g}^{-}(h)$. The
variation of $\Delta T_g$ as a function of $h$ is then given by a
power law $\Delta T_g \propto h^{-1/\nu_3}$. 2) the case of strong
interactions. When increasing $\epsilon$, equation ~(\ref{eq:36})
must be considered as leveling off when $\epsilon$ becomes larger
than $\epsilon_{max}$ which is a few tens of $\epsilon_c$, so that the
$T_g$ shift saturates at $\Delta T^+_g$. This is a direct consequence
of the model proposed here. Increasing indefinitely the
interaction energy does not change the $T_g$ shift. In the strong
interaction limit, the $T_g$ shift is equal to $\Delta T^+_g(h)$
and has the same power law dependence on $h$: $\Delta T_g \propto
h^{-1/\nu_3}$. In the crossover region, the $T_g$ shift is a
function of both $h$ and $\epsilon$. Since $\epsilon_c$ itself
depends on $h$, the $T_g$ shift is not exactly a power law of $h$
in this regime and the complete equation ~(\ref{eq:36}) must be
used to fit the data. When considering experimental results, one
needs thereby to check whether one deals with one of the two
limiting case, or whether the system is in the cross-over in
between. Let us now discuss  some
experimental data regarding the shift of $T_g$ on various
substrates and conditions within this framework.

Kawana and Jones \cite{jones2} have reported various measurements
of the glass transition temperature $T_g$ of polystyrene (PS)
films deposited on silicon wafers by spin coating, as a function
of the thickness $h$ of the film. $T_g$ is measured from thermal
expansivity curves obtained by ellipsometry. In these experiments,
$T_g$ is shifted downwards, which corresponds to a case where the
interactions between the polymer film and the substrate are weak.
The data were fitted using the empiric relation
\begin{equation}
T_{g}(h)=T_{g} \left[ 1- \left( \frac{A}{h} \right)^{\delta}
\right]                                  \label{eq:37}
\end{equation}
with a best fit value $\delta=1.1$. This value is indeed very close
to our prediction (eq(11)). However, the dispersion of
the data is too large  for discriminating precisely the value of
this exponent. The amplitude of the observed $T_g$ shifts
corresponds to a length $A=0.8$ nm. We shall compare this value to
our prediction for a film on a non-adsorbing substrate:
\begin{equation}
T_{g}(h)=T_{g} \left[ 1- \beta\left( \frac{a}{h}
\right)^{1/\nu_{3}} \right]
                              \label{eq:38}
\end{equation}
with $\beta=-\mu F'(p_{c}^{3D})/N_{c}^{1/2-1/3\nu}$. The values
$\mu \approx 0.5$, $F'(p_{c}^{3D}\approx 0.11) \approx -4$, $N_{c}
\approx 100$, $\nu \approx 0.88$, $a \approx 0.6$ nm give $A=a
\beta^{\nu} \approx 0.7$ nm, which is in remarkable agreement with
the measured value. Our prediction is thereby in quantitative
agreement with Kawana and Jones data, and we conclude that their
experiments correspond indeed to the case of a non-interacting
film, for which the effect of the interactions with the substrate
is negligible. We retain here that the typical $T_g$ shift for a
non-interacting film of thickness $h=10$ nm is about $20$ K (for a
bulk $T_g$ of the order 375 K), or about $10$ K at $h=20$ nm.

Hartmann et al \cite{kremer2} have reported dielectric measurements in isotactic
poly(methyl methacrylate) (i-PMMA) deposited on aluminium
electrodes evaporated on glass substrates, which show a decrease of
$T_g$ as $h$ decreases. The shift is of the order $10$ K for $h=20$
nm, which in this case also coincides with the value mentioned
above for a negligibly interacting film (with here $T_{g}$ of the
order $330$ K). We deduce thereby that in that case the interactions were
indeed negligible. In another series
of experiments, they have also performed ellipsometric measurements
in i-PMMA deposited on silica wafers, which show an
\emph{increase} of $T_g$ as $h$ decreases. This apparent
contradiction is interpreted as due to a difference in the
polymer-substrate interaction strength, because the experiments
have been performed on different substrates. Hartmann et al
interpreted these measurements in terms of a model by Herminghaus
 \cite{herminghaus}, which gives an expression for the $T_g$ shifts as a
function of the ratio $\sigma /E$ ($\sigma$ the surface tension,
$E$ the Young's modulus of the polymer at $T_g$). However, within Herminghaus
model, only decreases of $T_g$ are predicted, disregarding of
the polymer-substrate interaction. Changing the sign of the
$T_g$ shift as predicted by Herminghaus for interpreting
Hartmann and co-workers experiments is thereby
an unsubstantiated extension of this model \cite{long3}.
On Fig.~\ref{tgfig6} we plotted Hartman and al's results
obtained  by dielectric measurements, for two different molecular weights.
The straight line
represents the extrapolation of the data according to
equation ~(\ref{eq:38}), with a slope $-1/\nu_{3}=-1.136$. The data are
compatible with this slope, even though they are also not precise
enough to test this exponent. The ordinate at $\log h=0$ gives the
coefficient $A \approx 1$ nm.
Equation ~(\ref{eq:36}) of the present paper provides an explicit dependence of
the $T_g$ shift as a function of the interaction strength
$\epsilon$
in which $\Delta T_{g}^{-}(h)$ is the negative shift obtained for
a non-interacting film. As discussed above, the crossover
value $\epsilon_c$ contains
itself a $h$ dependence (cf equation ~(\ref{eq:35})) and in this regime, the
$T_g$ shift is not precisely a power law.
Nevertheless, if the $h$-dependence in $\epsilon_c$ is neglected
in equation ~(\ref{eq:36}), the intersection of
the solid line
in Fig.~\ref{fig7} with the ordinate axis provides the
pre-factor of equation ~(\ref{eq:36}), i.e. $\log
((\epsilon/\epsilon_{c})^{1/\gamma_{2}}-1) \approx 0.7$, or
$(\epsilon/\epsilon_c)^{1/\gamma_{2}} \approx 6$, that is
$\epsilon \approx 75 \epsilon_c$ -assuming $\log A=0$, i.e. $A \approx
1$ nm. From that we deduce that these
experiments have been performed in the regime of strong interactions.
In Fig.~\ref{fig7}) we have indicated the slope $-1.13$ which fits
reasonably well the data.
Fryer and co-workers  \cite{nealey2} have measured the $T_g$
shift of PMMA films of
various thicknesses deposited on silicon oxide, using a
thermal probe technique. Here also, positive shifts have been reported,
which correspond thereby to non-negligible interactions between the
polymer film and the substrate. However, the positive shift here
is not as large as in the previously discussed case. In
Fig.~\ref{fig8}), the ordinate at $\log h=0$ gives $\log
((\epsilon/\epsilon_{c})^{1/\gamma_{2}}-1) \approx -0.2$, that is
$\epsilon \approx 3 \epsilon_c$ (with $A \approx 1$ nm). We are
thus here in an intermediate regime, where the ratio
$\epsilon/\epsilon_c$ is not very large with respect to one.
Therefore the variation of $\Delta T_g$ versus $h$ should be better
described using the exact equation ~(\ref{eq:36}). However, in the above
discussion, the data have been extrapolated using the power law
$h^{-1.136}$.  Indeed, the data are not precise enough and the $h$ range is
too small to discriminate the strong adsorption limit and the cross-over
regime, as far as the variation of $T_g$ versus $h$ is concerned. The amplitude
of the shift, however, allows for this discrimination.
In another series of experiments,
Fryer et al \cite{nealey1} have measured the $T_g$ shift in PS and PMMA
films deposited on substrates with tuned interfacial energy. This
was achieved by using as substrates octadecyltrichlorosilane (OTS)
films deposited on silicon wafers and exposed to various doses of
X-Rays. The authors claim that, at a given thickness, the $T_g$ shift
 varies linearly with the interfacial energy $\sigma$, which at
 first sight may seem in apparent contradiction with our prediction
(equation ~(\ref{eq:36})). However, we stress that the
 adsorption energy  $\epsilon$ and the
 interfacial energy $\sigma$ are two different quantities,
 and that the relation between these two parameters is not
straightforward \cite{nealey1,isr}. Nevertheless, the data
 published by  Fryer and co-workers offer the opportunity to estimate
 the ratio $\epsilon/\epsilon_c$  using our model 
 in the different cases considered in \cite{nealey1}.
\begin{table}
\caption{\label{tab:table1}Values of the $T_g$ shift for a
thickness $h=20$ nm, for PS and PMMA films deposited on substrates
with various interfacial energies, from the data in Fryer et al
(2001). For PS: $T_{g}=375$ K; for PMMA: $T_{g}=392$ K. The values
given for $\epsilon/\epsilon_c$ are based on a shift of 10 K for
$h=20$ nm, and the $h$ variation in $\epsilon_c$ is neglected.}
\begin{ruledtabular}
\begin{tabular}{lccr}
Polymer&$\sigma$ (mJ/m$^2$)&$\Delta T_g$&$\epsilon/\epsilon_c$\\
\hline
PS & 0.8 & -8 & 0.02\\
PS & 1.7 & -4 & 0.3\\
PS & 6.5 & +30 & 27\\
\hline
PMMA & 0.5 & -8 & 0.02\\
PMMA & 1.3 & -5 & 0.2\\
PMMA & 3.3 & +11 & 6\\
\end{tabular}
\end{ruledtabular}
\end{table}
For each polymer (PS or PMMA), a series of results
obtained for 3 different thicknesses are presented, for 3 values
of the interfacial energy (obtained by an X-Ray irradiation with 3
different doses). The thickness-dependent $T_g$ shift is either
negative or positive, according to the interfacial energy. The
value $\sigma_c$ at the crossover is obtained as the point where
$\Delta T_g = 0$ on each curve $\Delta T_g$ versus $\sigma$, for each
given thickness $h$. It is observed that all curves (obtained for
various $h$ values) intersect at a point of coordinates $\Delta
T_g =0$, $\sigma \approx 2$mJ/m$^2$. In terms of interfacial
energy $\sigma$, the crossover should therefore correspond roughly
to the value $\sigma_{c} \approx 2 mJ/m^2$. Note however that the surface 
tension $\sigma$ is not 
the relevant parameter for describing the strength 
of the interaction as far as the $T_g$ shift is concerned: the 
relevant paremeter is the adsorption energy per monomer. More 
precisely the relation between the $T_g$ shift and $\sigma$ depends 
on the considered substrate and polymeric liquid. We expect only the 
relation between $Delta T_g$ and the adsorption energy $\epsilon$ to 
be universal as given by equation (39). 
Using equation  ~(\ref{eq:36}),
and by taking $\Delta T^-_g(h=20nm) \approx -10 K$, one can deduce
the ratio $\epsilon/\epsilon_c$. The results are summarised in Table I.
One can see there that the regime of strong interaction
is reached when $\epsilon/\epsilon_c \approx 27$. Note that the $T_g$ shift of 
$30$ $K$ for a film $20 nm$ thick is in the plateau regime. Thereby, our 
estimate here of $\epsilon/\epsilon_c$ is an underestimate, and the 
real adsorption energy might be higher. 

\section{\textbf{Conclusion}}

In conclusion, we have proposed here a model for calculating
the $T_g$ shift of thin polymer films, as a function of the
adsorption energy between the polymer and the substrate. We show also 
how to calculate the corresponding viscosity of the film using the
bulk WLF law with the corresponding temperature shift. Our prediction could
be tested directly by pulling a tip immerged in the film, such as an
AFM tip, or a probe pulled using an optical tweezer for instance. 
As a consequence of this model, we expect the viscosity of a suspended film 
to be strongly anisotropic. For instance, the viscosity measured by shearing the 
film between two susbtrates (without interaction) should be much larger than that 
 measured by pulling the probe as discussed in this paper. 
Such experiments could be performed in the two limiting
cases of non-interacting films and strongly interacting films, and 
also in the cross-over regime
of intermediate adsorption energy. However, 
testing our general relation quantitatively 
would require a precise knowledge
of the interaction between the monomers and the substrate.
Finally, our prediction of long ranged changes of viscoelastic
properties should be of importance for explaining the reinforcement
of filled materials such as filled elastomers \cite{landel,edwards,berriot}
and should help in tailoring their properties by adjusting the
adsorption energy between the particles and the matrix.

\section{\textbf{Figures}}
\pagebreak
\section*{}
\begin{figure}
\includegraphics{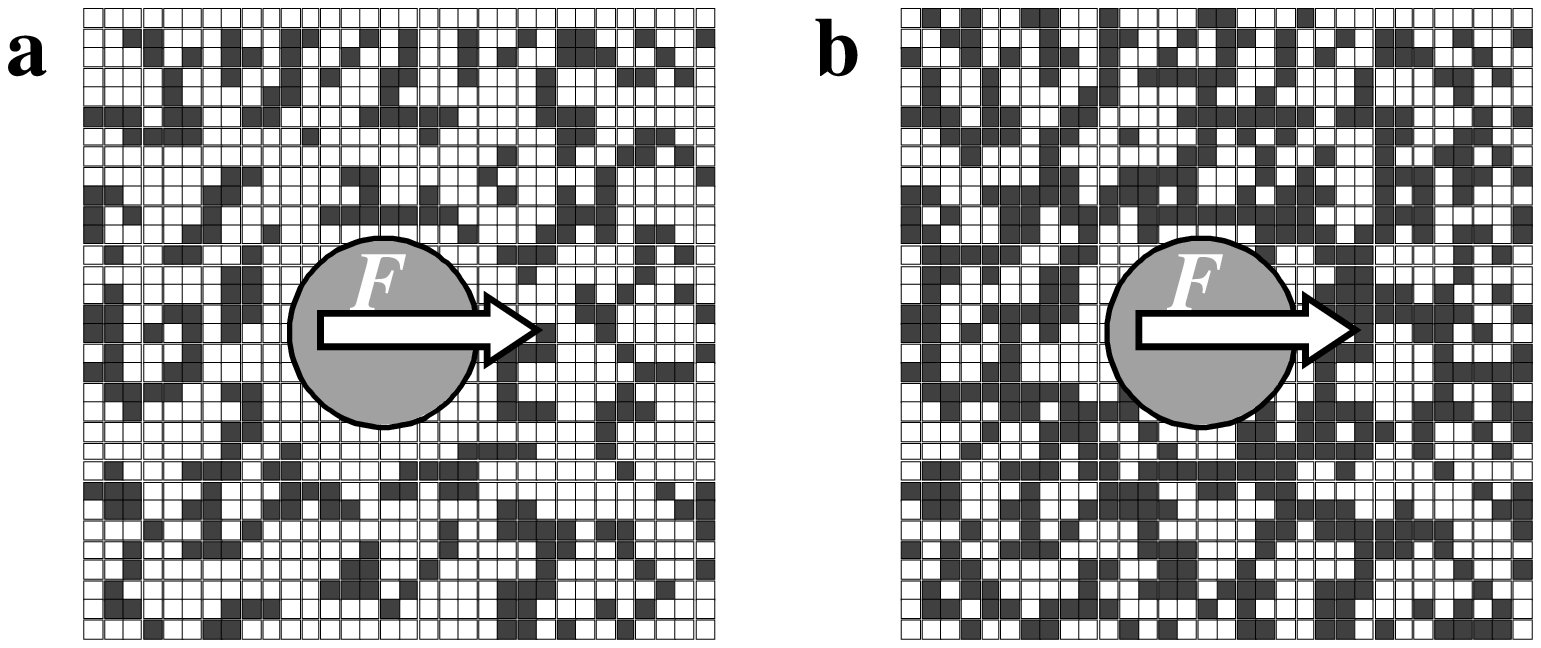}
\caption{\label{fig1}
We represent here a body floating in a thin suspended film, seen
from above. The body can be thought of emerging from the film on both interfaces.
We consider in Fig. 1.a a film at the bulk $T_g$. Since
the film is suspended, the slow subunits in Fig. 1.a, which do not percolate in the
direction parallel to the film, can move freely in the film to open the way to the
probe: the viscosity is smaller than in the bulk. By lowering the temperature (b), the
slow subunits percolate in the direction parallel to the film. They now control the
viscosity of the film as measured by the motion of the probe.}
\end{figure}
\pagebreak

\section*{}
\begin{figure}
\includegraphics{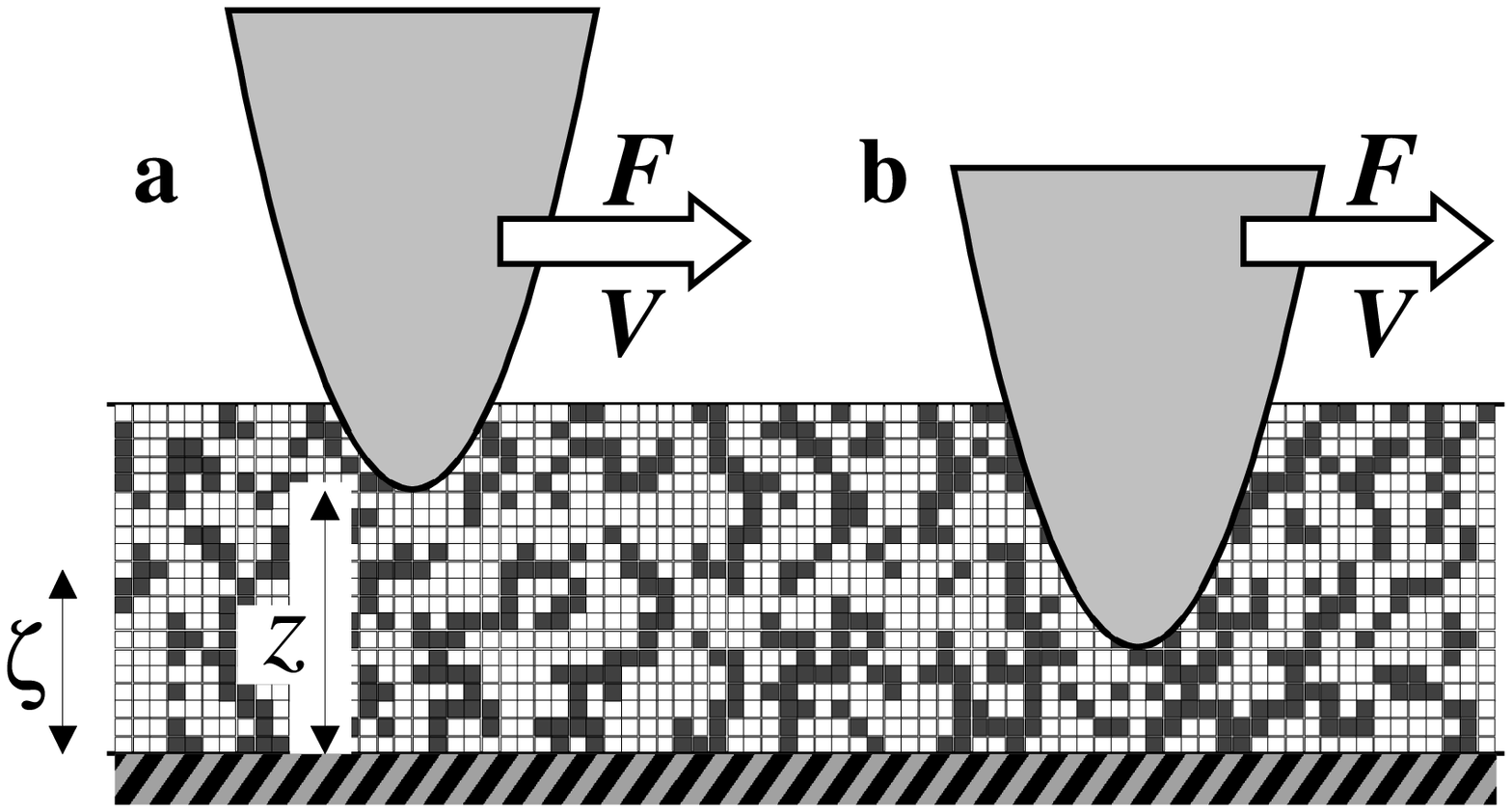}
\caption{\label{fig2} We consider here a film at a temperature $T$ higher than the bulk
$T_g$. In Fig.2.a, the tip is at a distance larger than the size $\zeta$ of
the slow aggregates. The slow aggregates can move freely around it, and the viscosity
is smaller than the viscosity at $T_g$. In Fig.2.b, the tip is closer to the substrate.
Moving parallel to the film requires deforming the slow aggregates: the viscosity is
larger than that at the bulk $T_g$.}
\end{figure}

\pagebreak
\section*{}
\begin{figure}
\includegraphics{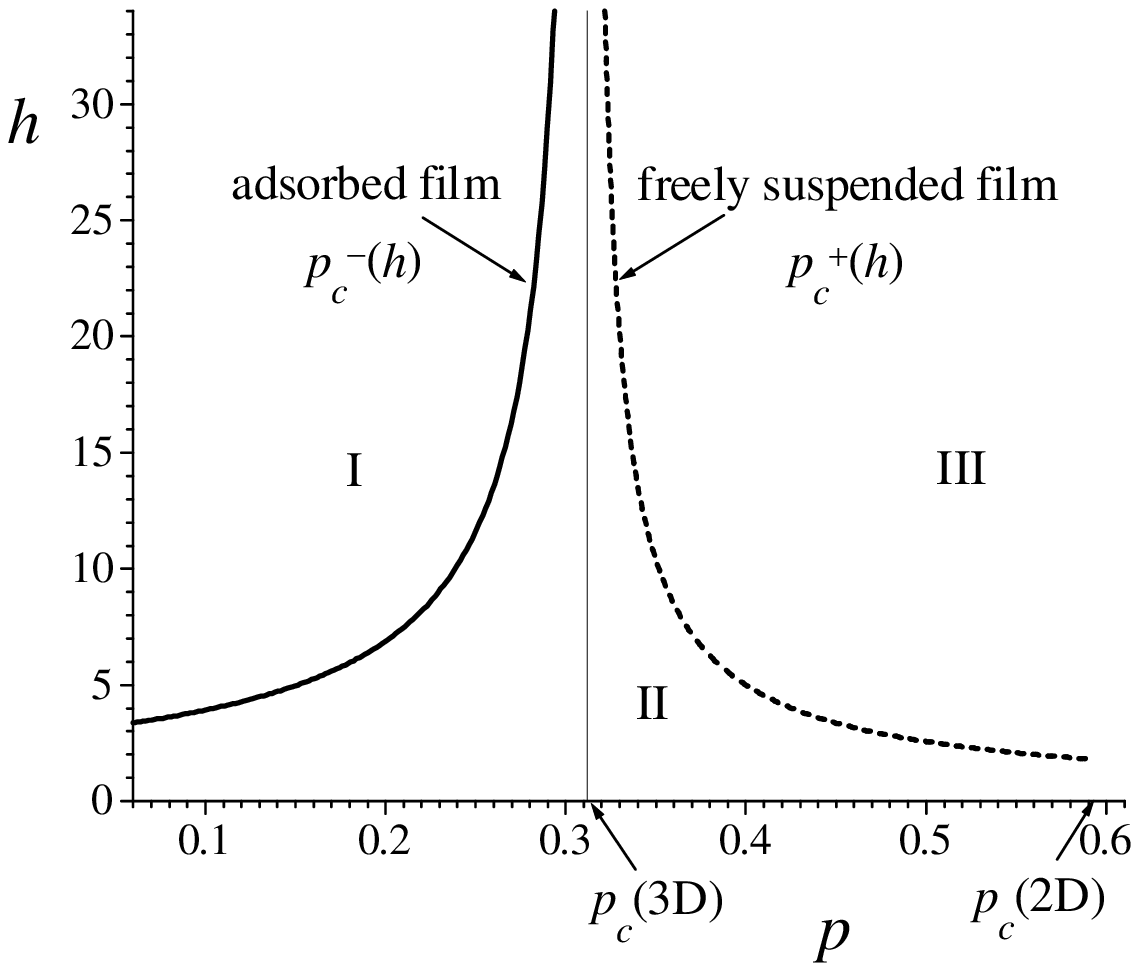}
\caption{\label{fig3} $p^+_c(h)$ and $p^-_c(h)$ as a function of $h$. 
Equivalently, this graph provides
the 3D correlation length of the percolation problem -within a prefactor-, as a function of $p$.
A freely suspended film of thickness $h$ is glassy at temperature $T$ if $p$
(related to $T$ by by equation
(8)) is larger than $p^+_c(h)$. A film with strong interactions with
the substrate is glassy if $p$ is larger than $p^-_c(h)$. Thereby, at any temperature
lower than $T_g+\Delta T^-_g$ a thin film is in the glassy state, while
at any temperature higher than $T_g+\Delta T^+_g$ the film is in the liquid
state. The cross-over between low interactions and strong interactions corresponds
to the region $p^-_c(h) \le p \le p^+_c(h)$.
}
\end{figure}

\pagebreak
\section*{}
\begin{figure}
\includegraphics{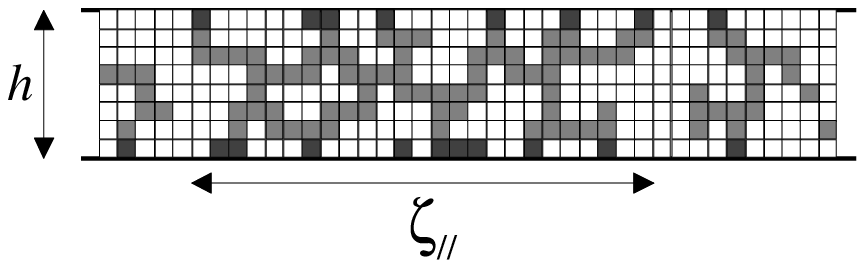}
\caption{\label{fig4}An aggregate of slow subunits in a thin film, in the
cross-over temperature range i.e. $T_g +\Delta T^-_g < T < T_g+\Delta T^+_g$.
The lateral extension of this aggregate is $\zeta_{\|}$, and the
number of subunits it contains is $M(h,p)$ }
\end{figure}

\pagebreak
\section*{}
\begin{figure}
\includegraphics{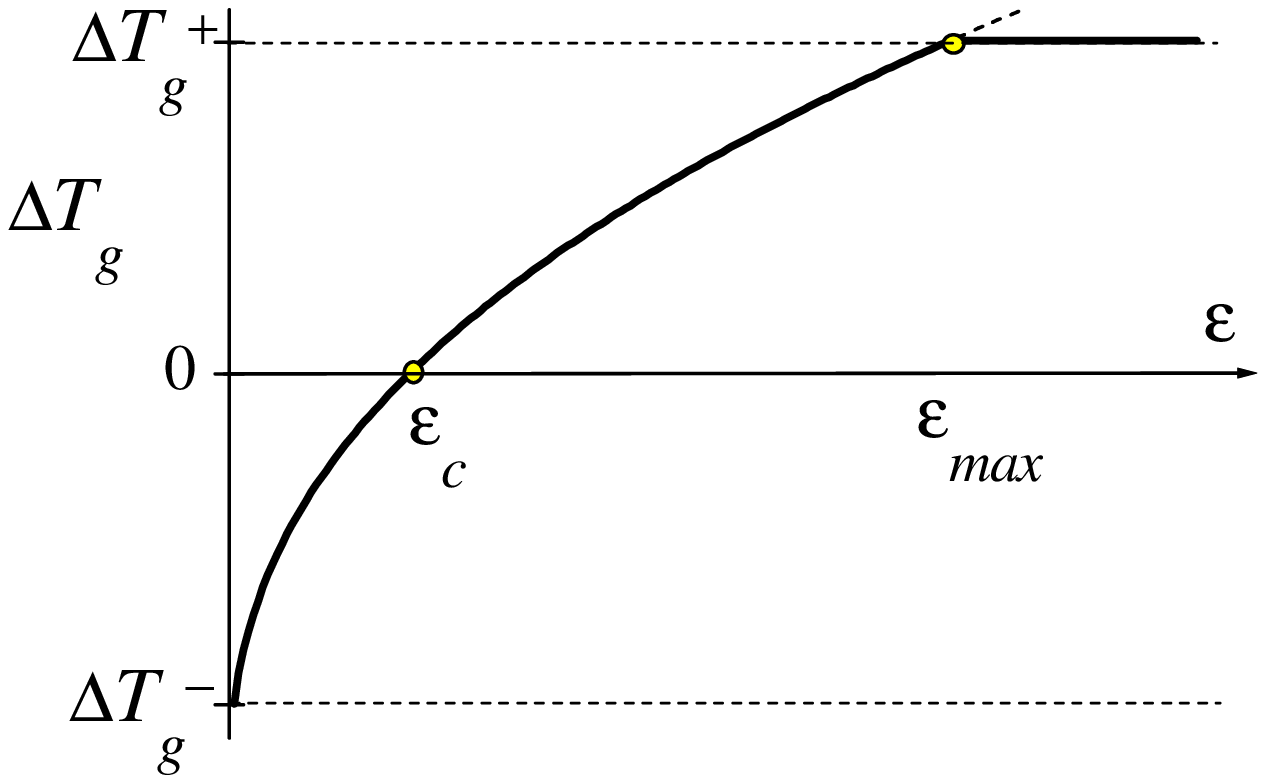}
\caption{\label{fig5} Variation of the $T_g$ shift of a thin film, as
a function of the interaction energy  between a monomer and
the substrate. The crossover between suspended films
and films with strong interactions with the substrate takes place at
$\epsilon = \epsilon_c \sim T$. }
\end{figure}
\pagebreak

\section*{}
\begin{figure}
\includegraphics{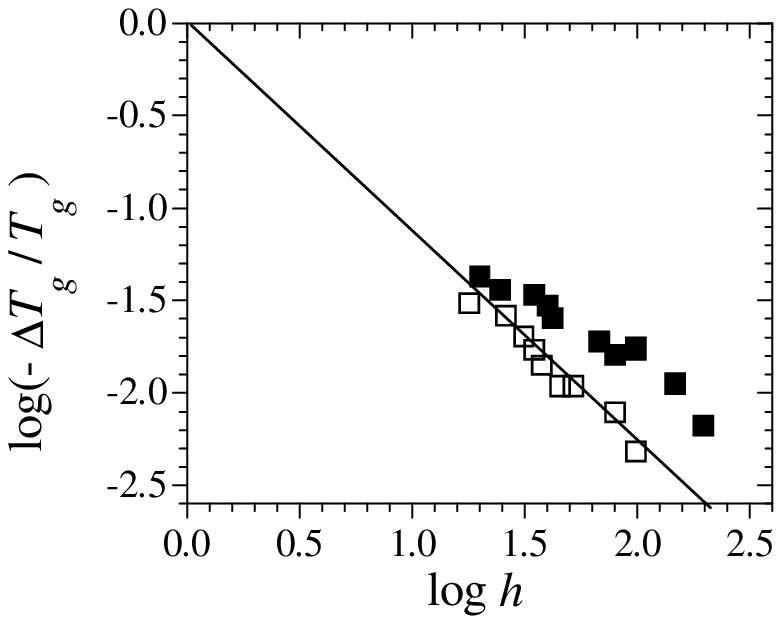}
\caption{\label{tgfig6}  Data from dielectric measurements in
i-PMMA films deposited on aluminium by Hartmann et al (2002).
$T_g$ decreases as $h$ decreases, so that $\Delta
T_{g}=T_{g}(h)-T_{g}$ is negative. Two different molecular weights 
are shown ($\square : M_{w}=44900$ g mol$^{-1}$, $T_{g}=330.6$ K,
$\blacksquare : M_{w}=164700$ g mol$^{-1}$, $T_{g}=331.2$ K). The
line has a slope -1.136. }
\end{figure}
\pagebreak

\section*{}
\begin{figure}
\includegraphics{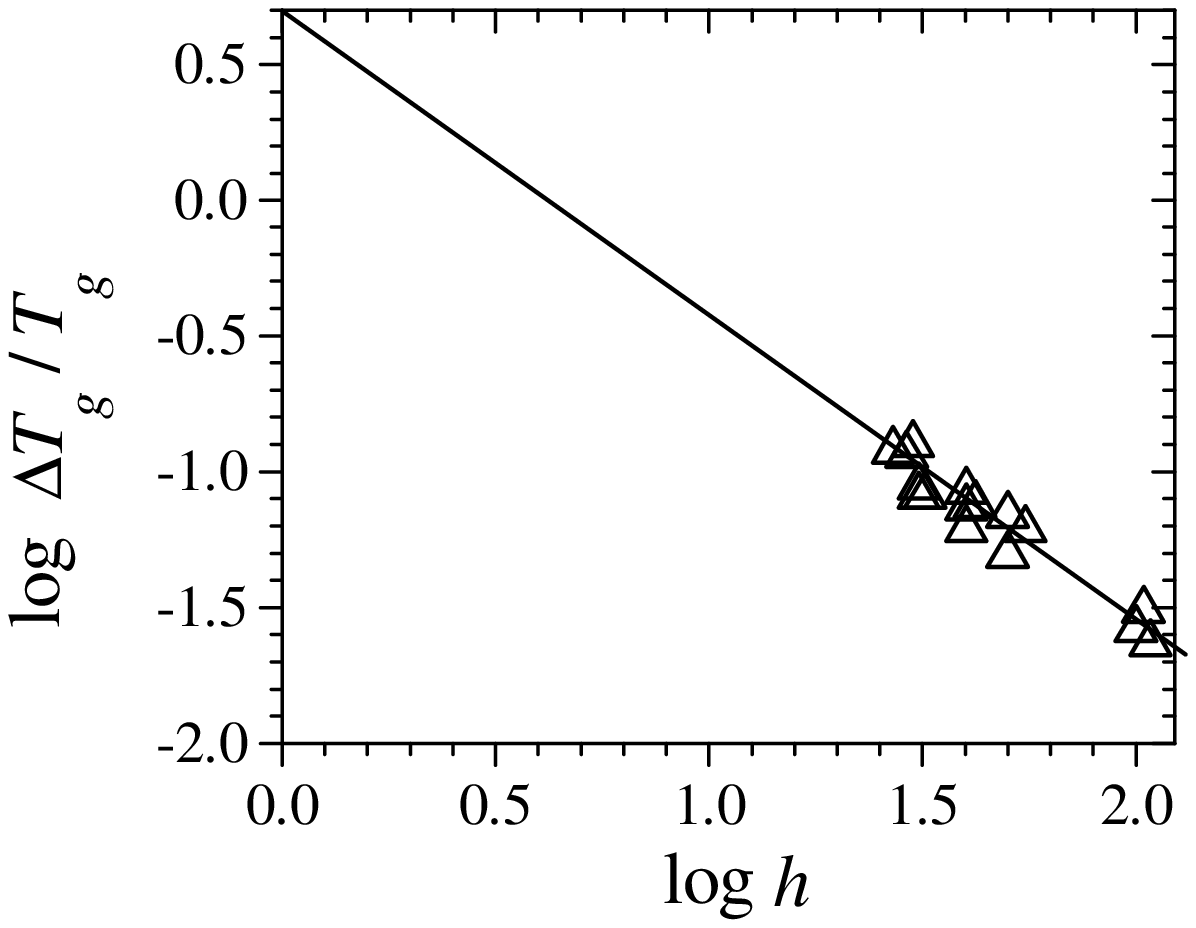}
\caption{\label{fig7} Data from ellipsometric measurements in
i-PMMA films deposited on silica wafers by Hartmann et al (2002).
$T_g$ increases as $h$ decreases, so that $\Delta T_g$ is positive
here. This case corresponds to strong interactions between the
polymer film and the substrate. ($M_{w}=44900$ g mol$^{-1}$). The
solid line has a slope -1.136.}
\end{figure}
\pagebreak

\section*{}
\begin{figure}
\includegraphics{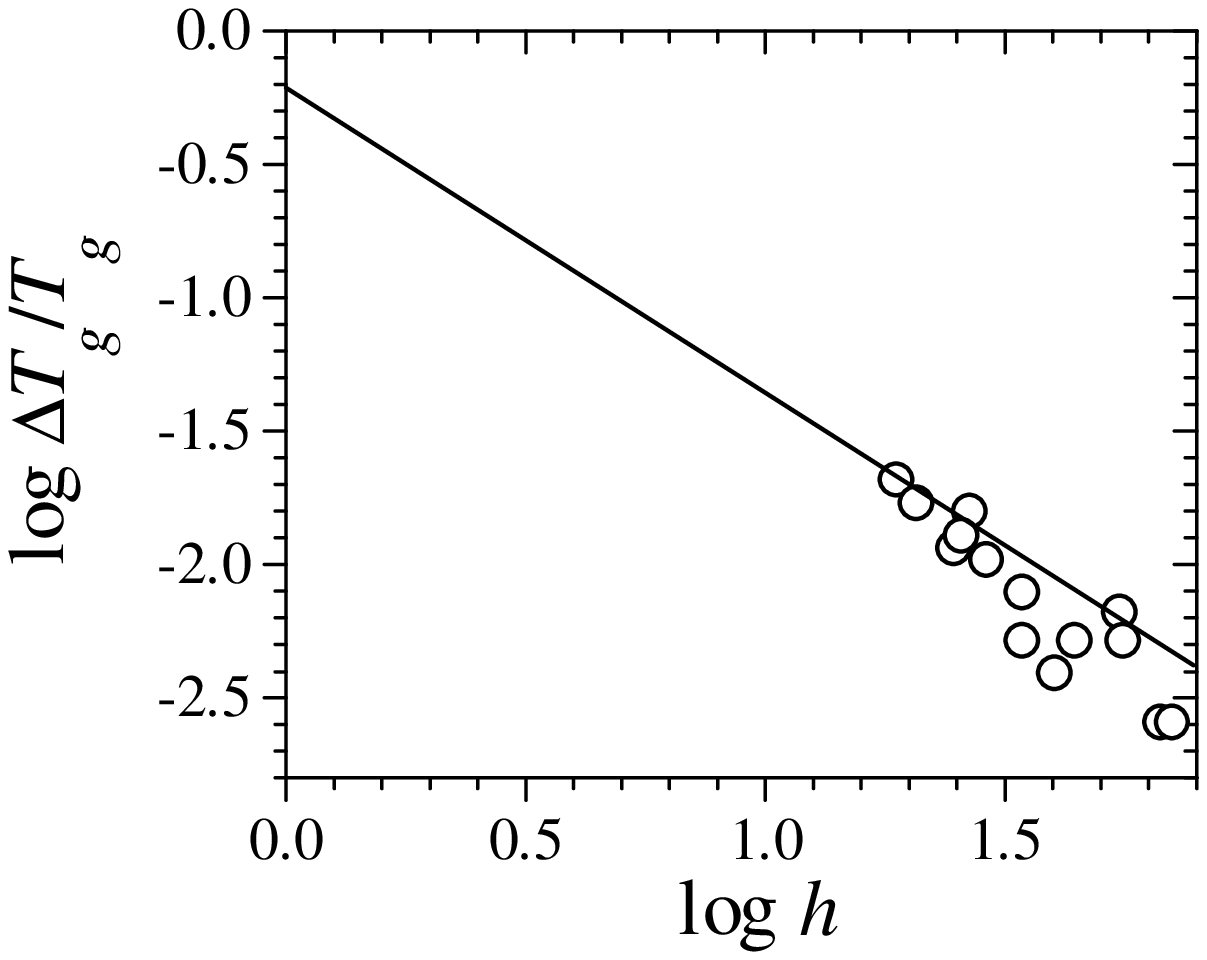}
\caption{\label{fig8} Data from thermal probe measurements in PMMA
films deposited on silicon oxide by Fryer et al (2000). $T_g$
increases as $h$ decreases, so that $\Delta T_g$ is positive here.
This case corresponds to intermediate adsorption energy between the polymer
film and the substrate. ($T_{g}=392K$). The solid line has a slope
-1.136.}
\end{figure}
\pagebreak

\end{document}